\documentclass[a4paper,twocolumn]{article}
\usepackage{epsfig}

\def\dprod{\mathop{\displaystyle \prod }}

\begin{document}
\title{Effects of Fast Presynaptic Noise in Attractor Neural Networks}
\author{J. M. Cortes$^{\dag \ddag}$, J. J. Torres$^{\dag}$, J. Marro$^{\dag}$, P. L. Garrido$^{\dag}$ and H. J. Kappen$^{\ddag}$ \\
$^{\dag}$Institute \textit{Carlos I} for Theoretical and Computational Physics, and \\
 Departamento de Electromagnetismo y F\'{\i}sica de la Materia,\\ 
University of Granada, E-18071 Granada, Spain.\\
$^{\ddag}$Department of Biophysics,  Radboud University of Nijmegen, \\
6525 EZ Nijmegen, The Netherlands}
\maketitle

{To appear in Neural Computation, 2005}

{Corresponding author: Jesus M. Cortes}

{mailto:jcortes@ugr.es}

\begin{abstract}
We study both analytically and numerically the effect of presynaptic noise on the transmission of information in attractor
neural networks. The noise occurs on a very short--time scale compared to
that for the neuron dynamics and it produces short--time synaptic depression. 
This is inspired in recent neurobiological
findings that show that synaptic strength may either increase or decrease on
a short--time scale depending on presynaptic activity. We thus describe a
mechanism by which fast presynaptic noise enhances the neural network
sensitivity to an external stimulus. The reason for this is that, in
general, the presynaptic noise induces nonequilibrium behavior and,
consequently, the space of fixed points is qualitatively modified in such a
way that the system can easily scape from the attractor. As a result, the
model shows, in addition to pattern recognition, class identification and
categorization, which may be relevant to the understanding of some of the
brain complex tasks.
\end{abstract}

\section{Introduction}

There is multiple converging evidence \cite{abb} that synapses determine the
complex processing of information in the brain. An aspect of this statement
is illustrated by attractor neural networks. These show that synapses can
efficiently store patterns that are afterwards retrieved with only partial
information on them. In addition to this long--time effect, artificial 
neural networks should contain some \textquotedblleft synaptic
noise\textquotedblright , however. That is, actual synapses 
exhibit short--time fluctuations, which seem to compete with other
mechanisms during the transmission of information, not to cause
unreliability but to ultimately determine a variety of computations \cite{allenPNAS,zadorJN}. In spite of some recent efforts, a full understanding of 
how the brain complex processes depend on such fast synaptic variations is 
lacking ---see below and \cite{abb}, for instance---. A specific
matter under discussion concerns the influence of short--time noise on the
fixed points and other details of the retrieval processes in attractor
neural networks \cite{bibitchkov}.

The observation that actual synapses endure short--time \textit{depression}
and/or \textit{facilitation} is likely to be relevant in this context. That
is, one may understand some observations by assuming that periods of
elevated presynaptic activity may cause either decrease or increase of the
neurotransmitter release and, consequently, that the postsynaptic response
will be either \textit{depressed} or \textit{facilitated} depending on presynaptic neural activity \cite{tsodyksNC,thom,abb}. Motivated
by the neurobiological findings, we report in this paper on effects of
presynaptic depressing noise on the functionality of a neural circuit. We study in
detail a network in which the neural activity evolves at random in time
regulated by a \textquotedblleft temperature\textquotedblright\ parameter.
In addition, the values assigned to the synaptic intensities by a \textit{learning} (e.g., Hebb's) rule are constantly perturbed with \textit{microscopic} fast noise. A new parameter is involved by this perturbation
that allows for a continuum transition from depression to normal operation.

As a main result, this paper illustrates that, in general, the addition of
fast synaptic noise induces a nonequilibrium condition. That is, our systems
cannot asymptotically reach equilibrium but tend to nonequilibrium steady
states whose features depend, even qualitatively, on dynamics \cite{marroB}.
This is interesting because, in practice, thermodynamic equilibrium is rare
in nature. Instead, the simplest conditions one observes are characterized
by a steady flux of energy or information, for instance. This makes the
model mathematically involved, e.g., there is no general framework such as
the powerful (equilibrium) Gibbs theory, which only applies to systems with
a single Kelvin temperature and a unique Hamiltonian. However, our system
still admits analytical treatment for some choices of its parameters and, in
other cases, we discovered the more intricate model behavior by a series of
computer simulations. We thus show that fast presynaptic depressing noise during
external stimulation may induce the system to scape from the attractor,
namely, the stability of fixed point solutions is dramatically modified.
More specifically, we show that, for certain versions of the system, the
solution destabilizes in such a way that computational tasks such as class
identification and categorization are favored. It is likely this is the
first time such a behavior is reported in an artificial neural network as a
consequence of biologically--motivated stochastic behavior of synapses.
Similar instabilities have been reported to occur in monkeys \cite{monkey}
and other animals \cite{animals}, and they are believed to be a main feature
in odor encoding \cite{olfact}, for instance.

\section{Definition of model}

Our interest is in a neural network in which a local stochastic dynamics is
constantly influenced by presynaptic \textit{noise}. Consider a set of $N$
binary neurons with configurations $\mathbf{S}\equiv \left\{ s_{i}=\pm
1;i=1,\ldots ,N\right\} .$ \footnote{Note that such binary neurons, although a 
crude simplification of nature, are
known to capture the essentials of cooperative phenomena, which is the focus
here. See, for instance \cite{AK90,torresNC}.}
Any two neurons are connected by synapses of intensity:
\footnote{For simplicity, we are neglecting here postsynaptic dependence of the 
stochastic perturbation. There is some claim that plasticity might operate on rapid 
time--scales on postsynaptic activity; see \cite{PAJN92}. However,
including $x_{ij}$ in (\ref{ws}) instead of $x_{j}$ would impede some of the
algebra in sections \ref{sect3} and \ref{sect4}.}
\begin{equation}
w_{ij}=\overline{w}_{ij}x_{j}\,\,\,\,\forall i,j.  \label{ws}
\end{equation}
Here, $\overline{w}_{ij}$ is fixed, namely, determined in a previous \textit{learning} process, and $x_{j}$ is a stochastic variable. This generalizes
the hypothesis in previous studies of attractor neural networks with noisy
synapses; see, for instance, \cite{S86,GM91,MTG99}. 
Once $\mathbf{W\equiv }\{\overline{w}_{ij}\}$ is given,
the state of the system at time $t$ is defined by setting 
$\mathbf{S}$ and $\mathbf{X}\equiv \{x_{i}\}.$ These evolve with time
---after the learning process which fixes $\mathbf{W}$--- via the familiar
Master Equation, namely,  
\begin{eqnarray}
\frac{\partial P_{t}(\mathbf{S},\mathbf{X})}{\partial t} =-P_{t}(\mathbf{S}
,\mathbf{X})\int_{\mathbf{{X^{\prime }}}}\sum_{\mathbf{{S^{\prime }}}}\,c[(
\mathbf{S},\mathbf{X})\rightarrow (\mathbf{{S^{\prime }}},\mathbf{{X^{\prime
}}})]  \nonumber \\
+\int_{\mathbf{{X^{\prime }}}}\sum_{\mathbf{{S^{\prime }}}}\,c[(\mathbf{{\
S^{\prime }}},\mathbf{{X^{\prime }}})\rightarrow (\mathbf{S},\mathbf{X}
)]P_{t}(\mathbf{{S^{\prime }}},\mathbf{{X^{\prime }}}).  \label{gme}
\end{eqnarray}
We further assume that the \textit{transition rate} or probability per unit
time of evolving from $(\mathbf{S},\mathbf{X})$ to $(\mathbf{S}^{\prime },
\mathbf{X}^{\prime })$ is 
\begin{eqnarray}
c[(\mathbf{S},\mathbf{X})\rightarrow (\mathbf{{S^{\prime }}},\mathbf{{\
X^{\prime }}})]=p\,c^{\mathbf{X}}[\mathbf{S}\rightarrow \mathbf{{S^{\prime }}
}]\delta (\mathbf{X}-\mathbf{{X^{\prime }}})  \nonumber \\ 
+(1-p)\,c^{\mathbf{S}}[\mathbf{X}
\rightarrow \mathbf{{X^{\prime }}}]\delta _{\mathbf{S},\mathbf{{S^{\prime }}}
}.  \label{gtr}
\end{eqnarray}
This choice \cite{garridoJSP,torresPRA} amounts to consider competing
mechanisms. That is, neurons $(\mathbf{S})$ evolve stochastically in time
under a noisy dynamics of synapses $(\mathbf{X}),$ the latter evolving 
$(1-p)/p$ times faster than the former. Depending on the value of $p,
$ three main classes may be defined \cite{marroB}:

\begin{enumerate}
\item For $p\in \left( 0,1\right) $ both the synaptic fluctuation and the
neuron activity occur on the same temporal scale. This case has already been
preliminary explored \cite{torresNC,cortes2T}.

\item The limiting case $p\rightarrow 1.$ This corresponds to neurons evolving in
the presence of a quenched synaptic configuration, i.e., $x_{i}$ is constant
and independent of $i.$ The \textit{Hopfield model} \cite{amari72,hopfield} belongs to 
this class in the simple case that $x_{j}=1,\forall j.$ 

\item The limiting case $p\rightarrow 0.$ The rest of this paper is devoted
to this class of systems.
\end{enumerate}

Our interest for the latter case is a consequence of the following facts. Firstly,
there is adiabatic elimination of fast variables for $p\rightarrow 0$ which
decouples the two dynamics \cite{garridoJSP,gardinerB}. Therefore,
some exact analytical treatment ---though not the complete solution--- is
then feasible. To be more specific, for $p\rightarrow 0,$ the neurons evolve
as in the presence of a steady distribution for $\mathbf{X.}$ If we write $P(
\mathbf{S},\mathbf{X})=P(\mathbf{X}|\mathbf{S})\,P(\mathbf{S}),$ where $P(
\mathbf{X}|\mathbf{S})$ stands for the conditional probability of $\mathbf{X}
$ given $\mathbf{S,}$ one obtains from (\ref{gme}) and (\ref{gtr}), after
rescaling time $tp\rightarrow t$  (technical details are worked out in \cite
{marroB}, for instance) that 
\begin{eqnarray}
\frac{\partial P_{t}(\mathbf{S})}{\partial t}=-P_{t}(\mathbf{S})\sum_{
\mathbf{{S^{\prime }}}}\bar{c}[\mathbf{S}\rightarrow \mathbf{{S^{\prime }}}
] \nonumber \\
+\sum_{\mathbf{{\ S^{\prime }}}}\bar{c}[\mathbf{{S^{\prime }}}\rightarrow 
\mathbf{S}]P_{t}(\mathbf{{S^{\prime }}}).  \label{geme}
\end{eqnarray}
Here, 
\begin{equation}
\bar{c}[\mathbf{S}\rightarrow \mathbf{{S^{\prime }}}]\equiv \int \mathrm{d}
\mathbf{X}\,P^{\mathrm{st}}(\mathbf{X}|\mathbf{S})\,c^{\mathbf{X}}[\mathbf{S}
\rightarrow \mathbf{{S^{\prime }}}],  \label{getr}
\end{equation}
and $P^{\mathrm{st}}(\mathbf{X}|\mathbf{S})$ is the stationary solution that
satisfies 
\begin{equation}
P^{\mathrm{st}}(\mathbf{X}|\mathbf{S})=\frac{\int \mathrm{d}\mathbf{{\
X^{\prime }}}\,c^{\mathbf{S}}[\mathbf{{X^{\prime }}}\rightarrow \mathbf{X}
]\,P^{\mathrm{st}}(\mathbf{{X^{\prime }}}|\mathbf{S})}{\int \mathrm{d}
\mathbf{{X^{\prime }}}\,c^{\mathbf{S}}[\mathbf{X}\rightarrow \mathbf{{\
X^{\prime }}}]}.  \label{pst}
\end{equation}
This formalism will allows us for modelling fast synaptic noise which, within 
the appropiate context, will induce sort of synaptic depression, as explained 
in detail in section \ref{sect4}.

The superposition (\ref{getr}) reflects the fact that activity is the result
of competition between different elementary mechanisms. That is, different
underlying dynamics, each associated to a different realization of the
stochasticity $\mathbf{X,}$ compete and, in the limit $p\rightarrow 0,$ an 
\textit{effective} rate results from combining \textbf{$c^{\mathbf{X}}[
\mathbf{S}\rightarrow \mathbf{\ {S^{\prime }}}]$ }with probability \textbf{$P^{\mathrm{st}}(\mathbf{X}|\mathbf{S})$} for varying $\mathbf{X.}$ Each of
the elementary dynamics tends to drive the system to a well-defined
equilibrium state. The competition will, however, impede equilibrium and, in
general, the system will asymptotically go towards a {\em nonequilibrium} steady
state \cite{marroB}. The question is if such a competition between synaptic
noise and neural activity, which induces nonequilibrium, is at the origin of
some of the computational strategies in neurobiological systems. Our study
below seems to indicate that this is a sensible issue. As a matter of fact,
we shall argue below that $p\rightarrow 0$ may be realistic \textit{a priori}
for appropriate choices of \textbf{$P^{\mathrm{st}}(\mathbf{X}|\mathbf{S}).$}

For the sake of simplicity, we shall be concerned in this paper with
sequential updating by means of single neuron or \textquotedblleft
spin--flip\textquotedblright\ dynamics. That is, the elementary dynamic step
will simply consist of local inversions $s_{i}\rightarrow -s_{i}$ induced by
a bath at temperature $T.$ The elementary rate $c^{\mathbf{X}}[\mathbf{S}
\rightarrow \mathbf{{S^{\prime }}}]$ then reduces to a single site rate that
one may write as $\Psi \lbrack u^{\mathbf{\ X}}(\mathbf{S},i)].$ Here, $u^{
\mathbf{X}}(\mathbf{S},i)\equiv 2T^{-1}s_{i}h_{i}^{\mathbf{X}}(\mathbf{S}),$
where $h_{i}^{\mathbf{X}}(\mathbf{S})=\sum_{j\neq i}\overline{w}
_{ij}x_{j}s_{j}$ is the net presynaptic current arriving to ---or local
field acting on--- the (postsynaptic) neuron $i.$ The function $\Psi (u)$ is
arbitrary except that, for simplicity, we shall assume $\Psi (u)=\exp
(-u)\Psi (-u),$ $\Psi (0)=1$ and $\Psi (\infty )=0$ \cite{marroB}. 
We shall report on the consequences of more complex dynamics in a
forthcomming paper \cite{cortesgs05}.

\section{\label{sect3}Effective local fields}

Let us define a function $H^{\mathrm{eff}}(\mathbf{S})$ through the
condition of detailed balance, namely,
\begin{equation}
\frac{{\bar{c}}[\mathbf{S}\rightarrow \mathbf{S}^{i}]}{\bar{c}[\mathbf{S}
^{i}\rightarrow \mathbf{S}]}=\exp \left\{ -\left[ H^{\mathrm{eff}}(\mathbf{S}
^{i})-H^{\mathrm{eff}}(\mathbf{S})\right] T^{-1}\right\} .  \label{Heff}
\end{equation}%
Here, $\mathbf{S}^{i}$ stands for $\mathbf{S}$ after flipping at $i,$ $
s_{i}\rightarrow -s_{i}.$ We further define the \textquotedblleft effective
local fields\textquotedblright\ $h_{i}^{\mathrm{eff}}(\mathbf{S})$ by means
of
\begin{eqnarray}
H^{\mathrm{eff}}(\mathbf{S})=-\frac{1}{2}\sum_{i}h_{i}^{\mathrm{eff}}(
\mathbf{S})\,s_{i}.  \label{eh}
\end{eqnarray}
Nothing guaranties that $H^{\mathrm{eff}}(\mathbf{S})$ and $%
h_{i}^{\mathrm{eff}}(\mathbf{S})$ have a simple expression and are therefore
analytically useful. This is because the superposition (\ref{getr}), unlike
its elements $\Psi (u^{\mathbf{\ X}}),$ does not satisfy detailed balance,
in general. In other words, our system has an essential nonequilibrium
character that prevents one from using Gibbs's statistical mechanics, which requires a unique
Hamiltonian. Instead, there is here one energy associated with each
realization of $\mathbf{X=}\{x_{i}\}.$ This is in addition to the fact that
the synaptic weights $w_{ij}$ in (\ref{ws}) may not be symmetric.

For some choices of both the rate $\Psi $ and the noise distribution $P^{
\mathrm{st}}(\mathbf{X}|\mathbf{S}),$ the function $H^{\mathrm{eff}}(\mathbf{
S})$ may be considered as a true effective Hamiltonian \cite
{garridoPRL,marroB}. This means that $H^{\mathrm{eff}}(\mathbf{S})$ then
generates the same nonequilibrium steady state than the stochastic
time--evolution equation which defines the system, i.e., equation (\ref{geme}), and that its coefficients have the proper symmetry of interactions. To be
more explicit, assume that \textbf{$P^{\mathrm{st}}$}$\left( \mathbf{\mathbf{
X}|\mathbf{S}}\right) $ factorizes according to
\begin{equation}
\mathbf{P^{\mathrm{st}}}\left( \mathbf{\mathbf{X}|\mathbf{S}}\right)
=\dprod\limits_{j}P\left( x_{j}\mathbf{|}s_{j}\right) ,  \label{prod}
\end{equation}
and that one also has the factorization
\begin{equation}
\bar{c}[\mathbf{S}\rightarrow \mathbf{S}^{i}]=\prod_{j\neq i}\int \mathrm{d}
x_{j}\,P(x_{j}|s_{j})\,\Psi (2T^{-1}s_{i}\overline{w}_{ij}x_{j}s_{j}).
\label{ceff}
\end{equation}
The former amounts to neglect some global dependence of the factors on $
\mathbf{S=}\left\{ s_{i}\right\} $ (see below), and the latter restricts the
possible choices for the rate function. Some familiar choices for this
function that satisfy detailed balance are: the one corresponding to the
Metropolis algorithm, i.e., $\Psi (u)=\min [1,\exp (-u)];$ the Glauber
case $\Psi (u)=[1+\exp (u)]^{-1};$ and $\Psi (u)=\exp (-u/2)$ \cite{marroB}. The latter fulfills $\Psi (u+v)=\Psi (u)\Psi (v)$ which is
required by (\ref{ceff}) \footnote{ In any case, the rate needs to be properly normalized. 
In computer simulations, it is customary
to divide $\Psi(u)$ by its maximum value. Therefore, the normalization happens to
depend on temperature and on the number of stored patterns. It follows that
this normalization is irrelevant for the properties of the steady state,
namely, it just rescales the time scale.}. It then ensues after some
algebra that
\begin{equation}
h_{i}^{\mathrm{eff}}=-T\sum_{j\neq i}\left[ \alpha _{ij}^{+}s_{j}+\alpha
_{ij}^{-}\right] ,  \label{gelf}
\end{equation}
with 
\begin{equation}
\alpha _{ij}^{\pm }\equiv \frac{1}{4}\ln \frac{{\bar{c}}(\beta _{ij};+)\,{
\bar{c}}(\pm \beta _{ij};-)}{{\bar{c}}(-\beta _{ij};\mp )\,{\bar{c}}(\mp
\beta _{ij};\pm )},  \label{alphas}
\end{equation}
where $\beta _{ij}\equiv 2T^{-1}\overline{w}_{ij},$ and 
\begin{equation}
\bar{c}(\beta _{ij};s_{j})=\int {d}x_{j}\,P(x_{j}|s_{j})\,\Psi (\beta
_{ij}x_{j}).
\end{equation}
This generalizes a case in the literature for random $\mathbf{S}$
--independent fluctuations \cite{garridoPRE, lacombaEURLETT, marroB}. In this
case, one has ${\bar{c}}(\pm \kappa ;+)={\bar{c}}(\pm \kappa ;-)$ and,
consequently, $\alpha _{ij}^{-}=0$ $\forall i,j.$ However, we here are
concerned with the case of $\mathbf{S}$--dependent disorder, which results
in a non--zero threshold, $\theta _{i}\equiv \sum_{j\neq i}\alpha
_{ij}^{-}\neq 0.$

In order to obtain a true effective Hamiltonian, the coefficients $\alpha
_{ij}^{\pm }$ in (\ref{gelf}) need to be symmetric. Once $\Psi (u)$ is
fixed, this depends on the choice for $P(x_{j}|s_{j}),$ i.e., on the fast noise
details. This is studied in the next section. Meanwhile, we remark that the
effective local fields $h_{i}^{\mathrm{eff}}$ defined above are very useful
in practice. That is, they may be computed ---at least numerically--- for
any rate and noise distribution. As far as $\Psi (u+v)=\Psi (u)\Psi (v)$ and 
$\mathbf{P^{\mathrm{st}}}\left( \mathbf{\mathbf{X}|\mathbf{S}}\right) $
factorizes,\footnote{\normalsize The factorization here does not need to be in products $P\left( x_{j}{|}s_{j}\right) $ as in (\ref{prod}). The same result (\ref{etr2}) holds for
the choice that we shall introduce in the next section, for instance.} it
follows an effective transition rate as
\begin{equation}
{\bar{c}}[\mathbf{S}\rightarrow \mathbf{S}^{i}]=\exp \left( -s_{i}h_{i}^{
\mathrm{eff}}/T\right) .  \label{etr2}
\end{equation}
This effective rate may then be used in computer simulation, and it may also
serve to be substituted in the relevant equations. Consider, for instance,
the \textit{overlaps} defined as the product of the current state with one
of the stored patterns:
\begin{equation}
m^{\nu }({\mathbf{S}})\equiv \frac{1}{N}\sum_{i}s_{i}\xi _{i}^{\nu }.
\label{OVER}
\end{equation}
Here, $\mathbf{\xi }^{\nu }=\{\xi _{i}^{\nu }=\pm 1,i=1,\ldots ,N\}$ are $M$
random patterns previously stored in the system, $\nu =1,\ldots ,M.$ After
using standard techniques \cite{hertzB,marroB}; see also \cite{amitAP}, it
follows from (\ref{geme}) that
\begin{equation}
\partial _{t}m^{\nu }=2N^{-1}\sum_{i}\xi _{i}^{\nu }\sinh \left( h_{i}^{
\mathrm{eff}}/T\right) -s_{i}\cosh \left( h_{i}^{\mathrm{eff}}/T\right) .
\label{kinetic}
\end{equation}
which is to be averaged over both thermal noise and pattern realizations.
Alternatively, one might perhaps obtain dynamic equations of type (\ref{kinetic}) by using Fokker-Planck like formalisms as, for instance, in \cite{brunel99}.

\section{\label{sect4}Types of synaptic noise}

The above discussion and, in particular, equations (\ref{gelf}) and (\ref
{alphas}), suggest that the system emergent properties will importantly
depend on the details of the synaptic noise\ $\mathbf{X.}$ We now work out
the equations in section \ref{sect3} for different hypothesis concerning the
stationary distribution (\ref{pst}).

Consider first (\ref{prod}) with the following specific choice:
\begin{equation}
P(x_{j}|s_{j})=\frac{1+s_{j}F_{j}}{2}\,\delta (x_{j}+\Phi )+\frac{
1-s_{j}F_{j}}{2}\,\delta (x_{j}-1).  \label{bimod}
\end{equation}
This corresponds to a simplification of the stochastic variable $x_{j}.$
That is, for $F_{j}=1$ $\,\forall j,$ the noise modifies $\overline{w}_{ij}$
by a factor $-\Phi $ when the presynaptic neuron is firing, $s_{j}=1,$ while
the learned synaptic intensity remains unchanged when the neuron is silent.
In general, $w_{ij}=-\overline{w}_{ij}\Phi $ with probability 
$\frac12$
$\left( 1+s_{j}F_{j}\right) .$ Here, $F_{j}$ stands for some information
concerning the presynaptic site $j$ such as, for instance, a local threshold
or $F_{j}=M^{-1}\sum_{\nu }\xi _{j}^{\nu }.$ 

Our interest for case (\ref{bimod}) is two fold, namely, it corresponds to
an exceptionally simple situation and it reduces our model to two known cases.
This becomes evident by looking at the resulting local fields:
\begin{equation}
h_{i}^{\mathrm{eff}}=\frac{1}{2}\sum_{j\neq i}\left[ \left( 1-\Phi \right)
s_{j}-\left( 1+\Phi \right) F_{j}\right] \overline{w}_{ij}.  \label{locelf}
\end{equation}
That is, exceptionally, symmetries here are such that the system is
described by a \textit{true} effective Hamiltonian. Furthermore, this
corresponds to the Hopfield model, except for a rescaling of temperature and
for the emergence of a threshold $\theta _{i}\equiv \sum_{j}\overline{w}
_{ij}F_{j}$ \cite{hertzB}. On the other hand, it also follows that,
concerning stationary properties, the resulting effective Hamiltonian (\ref
{eh}) reproduces the model as in \cite{bibitchkov}. In fact, this would
correspond in our notation to $h_{i}^{\mathrm{eff}}=\frac{1}{2}\sum_{j\neq i}
\overline{w}_{ij}s_{j}x_{j}^{\infty },$ where $x_{j}^{\infty }$ stands for
the stationary solution of certain dynamic equation for $x_{j}.$ The
conclusion is that (except perhaps concerning dynamics, which is something
worth to be investigated) the fast noise according to (\ref{prod}) with (\ref
{bimod}) does not imply any surprising behavior. In any case, this choice of
noise illustrates the utility of the effective--field concept as defined
above.

Our interest here is in modeling the noise consistent with the observation
of short-time synaptic depression \cite{tsodyksNC,torresNC}. In fact, the
case (\ref{bimod}) in some way mimics that increasing the mean firing rate 
results in decreasing the synaptic weight. With the same motivation, a more
intriguing behavior ensues by assuming, instead of (\ref{prod}), the
factorization 
\begin{equation}
P^{\mathrm{st}}(\mathbf{X}|\mathbf{S})=\prod_{j}P(x_{j}|\mathbf{S})
\label{xp}
\end{equation}
with 
\begin{equation}
P(x_{j}|\mathbf{S})=\zeta \left( \vec{\mathbf{m}}\right) \mathrm{\ }\delta
(x_{j}+\Phi )+\left[ 1-\zeta \left( \vec{\mathbf{m}}\right) \right] \mathrm{
\ }\delta (x_{j}-1).  \label{genbimod}
\end{equation}
Here, $\vec{\mathbf{m}}=\vec{\mathbf{m}}(\mathbf{S})\equiv \left( m^{1}(
\mathbf{S}),\ldots ,m^{M}(\mathbf{S})\right) $ is the $M$-dimensional
overlap vector, and $\zeta \left( \vec{\mathbf{m}}\right) $ stands for a
function of $\vec{\mathbf{m}}$ to be determined. The depression effect here
depends on the overlap vector which measures the net current arriving to
postsynaptic neurons. The non--local choice (\ref{xp})--(\ref{genbimod})
thus introduces non--trivial correlations between synaptic noise and neural
activity, which is not considered in (\ref{bimod}). Note that, therefore, we are not modelling here 
the synaptic depression dynamics in an explicity way as, for instance, in \cite{tsodyksNC}. 
Instead, equation (\ref{genbimod}) amounts to consider fast synaptic noise which naturally depresses the strengh of the synapses after repeated activity, namely, for a high value of $\zeta \left( \vec{\mathbf{m}}\right).$

Several further comments on the significance of (\ref{xp})-(\ref{genbimod}), which is here a
main hypothesis together with $p\rightarrow 0,$ are in order. We first mention that the
system time relaxation is typically orders of magnitude larger than the time
scale for the various synaptic fluctuations reported to account for the
observed high variability in the postsynaptic response of central neurons \cite{zadorJN}. On the other hand, these fluctuations seem to have different
sources such as, for instance, the stochasticity of the opening and closing
of the vesicles (S. Kilfiker, private communication), the stochasticity of
the postsynaptic receptor, which has its own several causes, variations of
the glutamate concentration in the synaptic cleft, and differences in the
potency released from different locations on the active zone of the synapses \cite{fssjn03}. Is this complex situation the one that we try to
capture by introducing the stochastic variable $x$ in (\ref{ws}) and subsequent
equations. It may be further noticed that the nature of this variable, which
is "microscopic" here, differs from the one in the case of familiar
phenomenological models. These often involve a "mesoscopic" variable, such
as the mean fraction of neurotransmitter, which results in a deterministic
situation, as in \cite{tsodyksNC}. The depression in our model rather
naturally follows from the coupling between the synaptic "noise" and the
neurons dynamics via the overlap functions. The final result is also
deterministic for $p\rightarrow 0$ but only, as one should perhaps expect, on the time scale for
the neurons. Finally, concerning also the reality of the model, it should be
clear that we are restricting ourselves here to fully connected networks
just for simplicity. However, we already studied similar systems with more
realistic topologies such as scale-free, small-world and diluted networks \cite{torresNEUCOM}, 
which suggests one to generalize the present study in
this sense.

It is to be remarked that our case (\ref{xp})-(\ref{genbimod}) also reduces
to the Hopfield model but only in the limit $\Phi \rightarrow -1$ for any $
\zeta \left( \vec{\mathbf{m}}\right) .$ Otherwise, the competition results
in a rather complex behavior. In particular, the noise distribution $P^{
\mathrm{st}}(\mathbf{X}|\mathbf{S})$ lacks with (\ref{genbimod}) the
factorization property which is required to have an effective Hamiltonian
with proper symmetry. Nevertheless, we may still write 
\begin{equation}
\frac{{\bar{c}}[\mathbf{S}\rightarrow \mathbf{S}^{i}]}{\bar{c}[\mathbf{S}
^{i}\rightarrow \mathbf{S}]}=\prod_{j\neq i}\frac{\int {d}x_{j}\,P(x_{j}|
\mathbf{S})\,\Psi (s_{i}x_{j}s_{j}\beta _{ij})}{\int {d}x_{j}\,P(x_{j}|
\mathbf{S}^{i})\,\Psi (-s_{i}x_{j}s_{j}\beta _{ij})}.  \label{db2}
\end{equation}
Then, using (\ref{genbimod}), we linearize around $
\overline{w}_{ij}=0,$ i.e., $\beta _{ij}=0$ for $T>0.$ This is a good
approximation for the Hebbian learning rule \cite{hebb} $\overline{w}
_{ij}=N^{-1}\sum_{\nu }\xi _{i}^{\nu }\xi _{j}^{\nu },$ which is the one we
use hereafter, as far as this rule only stores completely uncorrelated,
random patterns. In fact, fluctuations in this case are of order $\sqrt{M}/N$
for finite $M$ (or order $1/\sqrt{N}$ for finite $\alpha $) which tends
to vanish for a sufficiently large system, e.g., in the macroscopic
(thermodynamic) limit $N\rightarrow \infty .$ It then follows the effective
weights: 
\begin{equation}
w_{ij}^{\mathrm{eff}}=\left\{ 1-\frac{1+\Phi }{2}\left[ \zeta \left( \vec{
\mathbf{m}}\right) +\zeta \left( \vec{\mathbf{m}}^{\mathbf{i}}\right) \right]
\right\} \overline{w}_{ij},  \label{weff}
\end{equation}
where $\vec{\mathbf{m}}=\vec{\mathbf{m}}(\mathbf{S}),$ $\vec{\mathbf{m}}^{
\mathbf{i}}\equiv \vec{\mathbf{m}}(\mathbf{S}^{\mathbf{i}})=\vec{\mathbf{m}}
-2s_{i}\vec{\xi}_{i}/N,$ and $\vec{\xi}_{i}=\left( \xi _{i}^{1},\xi
_{i}^{2},...,\xi _{i}^{M}\right) $ is the binary $M$--dimensional stored
pattern. This shows how the noise modifies synaptic intensities. The
associated effective local fields are 
\begin{equation}
h_{i}^{\mathrm{eff}}=\sum_{j\neq i}w_{ij}^{\mathrm{eff}}s_{j}.
\label{gloelf}
\end{equation}
The condition to obtain a true effective Hamiltonian, i.e., proper symmetry
of (\ref{weff}) from this, is that $\vec{\mathbf{m}}^{\mathbf{i}}=\vec{
\mathbf{m}}-2s_{i}\vec{\xi}_{i}/N\simeq \vec{\mathbf{m}}.$ This is a good
approximation in the thermodynamic limit, $N\rightarrow \infty .$

\begin{figure}
\centerline{
\psfig{file=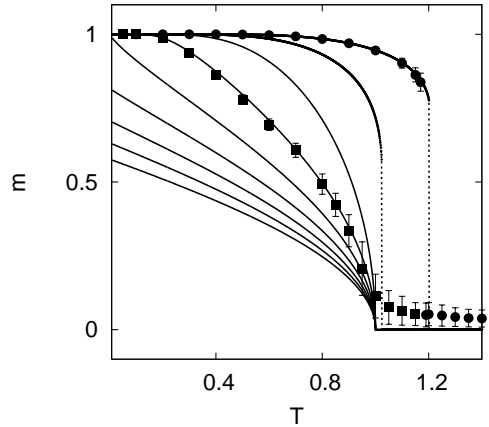,width=7cm}
}
\caption{{\footnotesize The steady overlap $m(T),$ as predicted by equation (\ref{stoe}), for different values of the noise parameter, namely, $\Phi
=-2.0,$ $-1.5,$ $-1.0,$ $-0.5,$ $0,$ $0.5,$ $1.0,$ $1.5,$ $2.0,$ from top to
bottom, respectively. ($\Phi =-1$ corresponds to the Hopfield case, as
explained in the main text.) The graphs depict second order phase
transitions (solid curves) and, for the most negative values of $\Phi,$ first order phase transitions (the discontinuities in these cases are
indicated by dashed lines). The symbols stand for Monte Carlo data
corresponding to a network with $N=1600$ neurons for $\Phi =-0.5$ (filled
squares) and $-2.0$ (filled circles).}}
\label{fig1}
\end{figure}

Otherwise, one may proceed with the dynamic equation (\ref{kinetic}) after
substituting (\ref{gloelf}), even though this is not then a true effective
Hamiltonian. One may follow the same procedure for the Hopfield case with
asymmetric synapses \cite{hertzB}, for instance. Further interest on the
concept of local effective fields as defined in section \ref{sect3} follows
from the fact that one may use quantities such as (\ref{gloelf}) to
importantly simplify a computer simulation, as it is made below.

To proceed further, we need to determine the probability $\zeta $ in (\ref
{genbimod}). In order to model activity--dependent mechanisms acting on the
synapses, $\zeta \mathbf{\ }$should be an increasing function of the net
presynaptic current or field. In fact, $\zeta \left( \vec{\mathbf{m}}\right) 
$ simply needs to depend on the overlaps, besides to preserve the $\pm 1$
symmetry. A simple choice with these requirements is 
\begin{equation}
\zeta \left( \vec{\mathbf{m}}\right) =\frac{1}{1+\alpha }\sum_{\nu }\left[
m^{\nu }\left( \mathbf{S}\right) \right] ^{2},  \label{psi}
\end{equation}
where $\alpha =M/N.$ We describe next the behavior that ensues from (\ref
{weff})--(\ref{psi}) as implied by the noise distribution (\ref{genbimod}). 

\section{Noise induced phase transitions}

\begin{figure}
\centerline{
\psfig{file=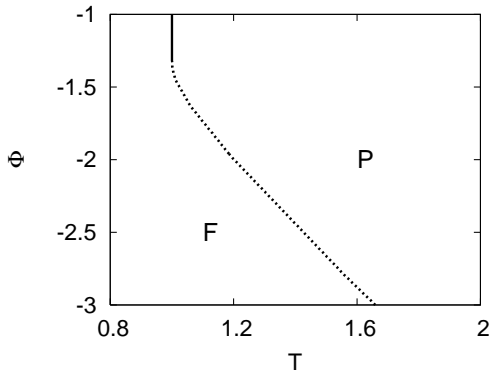,width=7cm}
}
\caption{ {\footnotesize Phase diagram depicting the transition temperature $T_{c}$ as a function of $T$ and $\Phi $. The solid (dashed) curve
corresponds to a second (first) order phase transition. The tricritical
point is at $(T_{c},\Phi _{c})=(1,-4/3).$ $F\,$and $P$ stand for the \textit{ferromagnetic--like} and \textit{paramagnetic--like} phases, respectively.
The best retrieval properties of our model system occur close to the
left--lower quarter of the graph.} }
\label{fig2}
\end{figure}

Let us first study the retrieval process in a system with a single stored
pattern, $M=1,$ when the neurons are acted on by the local fields (\ref
{gloelf}). One obtains from (\ref{etr2})--(\ref{kinetic}), after using the
simplifying (mean-field) assumption $\langle s_{i}\rangle \approx s_{i}$
that the steady solution corresponds to the overlap:
\begin{equation}
m=\tanh \left\{ T^{-1}m\left[ 1-(m)^{2}\left( 1+\Phi \right) \right]
\right\} ,  \label{stoe}
\end{equation}
$m\equiv m^{\nu =1},$ which preserves the symmetry $\pm 1.$ Local stability
of the solutions of this equation requires that
\begin{equation}
\left\vert m\right\vert >m_{c}(T)=\frac{1}{\sqrt{3}}\left( \frac{T_{c}-T}{
\Phi -\Phi _{c}}\right) ^{\frac{1}{2}}.  \label{stability}
\end{equation}

The behavior (\ref{stoe}) is illustrated in figure \ref{fig1} for several
values of $\Phi .$ This indicates a transition from a \textit{
ferromagnetic--like} phase, i.e., solutions $m\neq 0$ with associative
memory, to a \textit{paramagnetic--like} phase, $m=0.$ The transition is
continuous or second order only for $\Phi >\Phi _{c}=-4/3,$ and it then
follows a critical temperature $T_{c}=1.$ Figure \ref{fig2} shows the
tricritical point at $\left( T_{c},\Phi _{c}\right) $ and the general
dependence of the transition temperature with $\Phi .$

\begin{figure}
\centerline{
\psfig{file=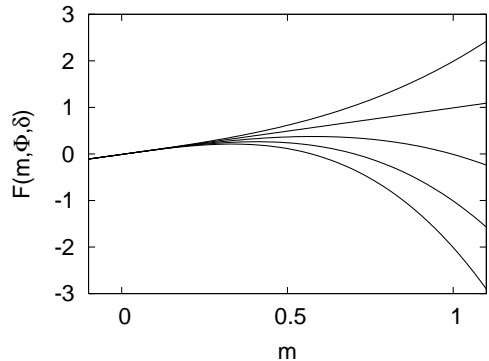,width=7cm}}
\caption{{\footnotesize The function $F$ as defined in (\ref{F}) for $\delta
=0$ and, from top to bottom, $\Phi =-2, -1, 0, 1$ and $2.$  The solution of (\ref{stoie}) becomes unstable so that the activity
will escape the attractor ($m=1$) for $F<0,$ which occurs for $\Phi >0
$ in this case.}}
\label{fig3}
\end{figure}

It is to be remarked that a discontinuous phase transition allows for a much
better performance of the retrieval process than a continuous one. This is
because the behavior is sharp just below the transition temperature in the
former case. Consequently, the above indicates that our model performs
better for large negative $\Phi ,$ $\Phi <-4/3.$

We also performed Monte Carlo simulations. These concern a network of $
N=1600 $ neurons acted on by the local fields (\ref{gloelf}) and evolving by
sequential updating via the effective rate (\ref{etr2}). Except for some
finite--size effects, figure \ref{fig1} shows a good agreement between our
simulations and the equations here; in fact, the computer simulations also
correspond to a mean--field description given that the fields (\ref{gloelf})
assume fully connected neurons.

\section{Sensitivity to the stimulus}

\begin{figure}
\psfig{file=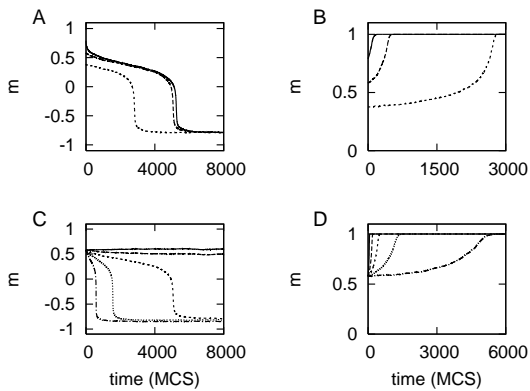,width=7.5cm}
\caption{{\footnotesize Time evolution of the overlap, as defined in (\ref{OVER}), between the current state and the stored pattern in Monte Carlo
simulations with 3600 neurons at $T=0.1.$ Each graph, for a given set of
values for $\left( \delta  ,\Phi \right)$ shows different curves
corresponding to evolutions starting with different initial states. The two
top graphs are for $\delta =0.3$ and $\Phi =1$ (graphs A and C) and $\Phi
=-1$ (graphs B and D), the latter corresponding to the Hopfield case lacking the
fast noise. This shows the important effect noise has on the network
sensitivity to external stimuli. The two bottom graphs illustrate the same
for a fixed initial distance from the attractor as one varies the external
stimulation, namely, for $\delta =0.1, 0.2, 0.3, 0.4$ and $0.5$ from top
to bottom.}}
\label{fig4}
\end{figure}

As shown above, a noise distribution such as (\ref{genbimod}) may model
activity-dependent processes reminiscent of short-time synaptic depression.
In this section, we study the consequences of this type of fast noise on the
retrieval dynamics under external stimulation. More specifically, our aim is
to check the resulting sensitivity of the network to external inputs. A high
degree of sensibility will facilitate the response to changing stimuli. This
is an important feature of neurobiological systems which continuously adapt
and quickly respond to varying stimuli from the environment.

\begin{figure}
\centerline{
\psfig{file=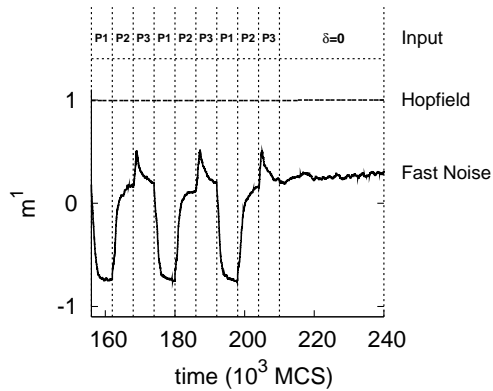,width=7cm}
}
\caption{{\footnotesize Time evolution during a Monte Carlo simulation with $N=400$ neurons, $M=3$ 
correlated patterns (as defined in the main
text), and $T=0.1$. The system in this case was let to relaxe to the
steady sate, and then perturbed by the stimulus $-\delta \mathbf{\xi }^{\nu
},$ $\delta =0.3,$ with $\nu =1$  for a short time interval, and then
with $\nu =2,$ and so on. After suppresing the stimulus, the system is
again allowed to relaxe. The graphs show as a function of time, from top to
bottom, (i) the number of the pattern which is used as the stimulus at each
time interval; (ii) the resulting response of the network, measured as the
overlap of the current state with pattern $\nu =1,$ in the absence of
noise, i.e., the Hopfield case $\Phi =-1;$ (iii) the same for the relevant
noisy case $\Phi =1.$}}
\label{fig5}
\end{figure}

Consider first the case of one stored pattern, $M=1.$ A simple external
input may be simulated by adding to each local field a driving term $-\delta
\xi _{i},\ \forall i,$ with $0<\delta \ll 1$ \cite{bibitchkov}. A negative
drive in this case of a single pattern assures that the network activity may
go from the attractor, $\mathbf{\xi ,}$ to the \textquotedblleft
antipattern\textquotedblright , $-\mathbf{\xi }.$ It then follows the
stationary overlap:
\begin{equation}
m=\tanh [T^{-1}F(m,\Phi ,\delta )]  \label{stoie}
\end{equation}
with
\begin{equation}
F(m,\Phi ,\delta )\equiv m[1-(m)^{2}(1+\Phi )-\delta ].  \label{F}
\end{equation}

Figure \ref{fig3} shows this function for $\delta =0$ and varying $\Phi .$
This illustrates two different types of behavior, namely, (local) stability $
(F>0)$ and instability $(F<0)$ of the attractor, which corresponds to $m=1.$
That is, the noise induces instability, resulting in this case in switching
between the pattern and the antipattern. This is confirmed in figure \ref
{fig4} by Monte Carlo simulations.

The simulations corresponds to a network of $N=3600$ neurons with one stored
pattern, $M=1.$ This evolves from different initial states, corresponding to
different distances to the attractor, under an external stimulus $-\delta 
\mathbf{\xi }^{1}$ for different values of $\delta .$ The two left graphs in
figure \ref{fig4} show several independent time evolutions for the model
with fast noise, namely, for $\Phi =1;$ the two graphs to the right are for
the Hopfield case lacking the noise $(\Phi =-1).$ These, and similar graphs
one may obtain for other parameter values, clearly demonstrate how the
network sensitivity to a simple external stimulus is qualitatively enhanced
by adding presynaptic noise to the system.

\begin{figure}
\centerline{\psfig{file=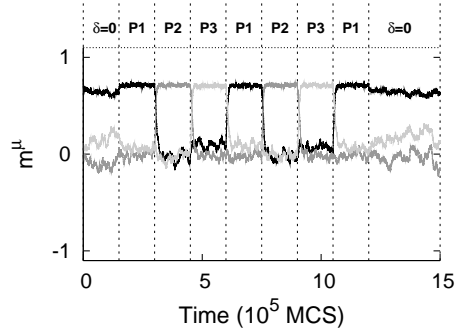,width=7cm}} 
\caption{{\footnotesize The same as in figure \ref{fig5} but for three stored patterns
that are orthogonal (instead of correlated). The stimulus is $+\delta \xi
^{\nu },$ $\delta =0.1,$ with $\nu =\nu \left( t\right) ,$ as indicated
at the top. The time evolution of the overlap $m^{\nu }$ is drawn with a
different color (black, dark-grey and light-grey, respectively) for each value 
of $\nu $ to illustrate that the system keeps jumping between the patterns in 
this case.}}
\label{fig6}
\end{figure}

Figures \ref{fig5} and \ref{fig6} illustrate a similar behavior in Monte
Carlo simulations with several stored patterns. Figure \ref{fig5} is for $M=3
$ \underline{correlated} patterns with 
mutual overlaps $|m^{\nu ,\mu }|\equiv
|1/N\sum_{i}\xi _{i}^{\nu }\xi _{i}^{\mu }|=1/3$ and $|\langle \xi _{i}^{\nu }\rangle|=1/3.$ 
More specifically, each
pattern consits of three equal initially white (silent neurons) horizontal stripes, with one
of them black colored (firing neurons) located in a different position for each pattern.
The system in this case begins with the first pattern as initial condition and,
to avoid dependence on this choice, it is let to relax for 3x10$^{4}$ Monte
Carlo steps (MCS). It is then perturbed by a drive $-\delta \mathbf{\xi }
^{\nu },$ where the stimulus $\nu $ changes $(\nu =1,2,3,1,...)$ every 6x10$
^{3}$ MCS. The top graph shows the network response in the Hopfield case.
There is no visible structure of this signal in the absence of fast noise as
far as $\delta \ll 1.$ As a matter of fact, the depth of the basins of
attraction are large enough in the Hopfield model, to prevent any move for
small $\delta ,$ except when approaching a critical point ($T_{c}=1$), where
fluctuations diverge. The bottom graph depicts a qualitatively different
situation for $\Phi =1.$ That is, adding fast noise in general destabilizes
the fixed point for the interesting case of small $\delta $ far from
criticality.

Figure \ref{fig6} confirms the above for \underline{uncorrelated} patterns, e.g. 
$m^{\nu ,\mu }\approx \delta^{\nu,\mu}$ and $\langle \xi_{i}^{\nu }\rangle \approx 0.$
That is, this shows the response of the network in a similar simulation with
400 neurons at $T=0.1$ for $M=3$ random, othogonal patterns. The initial
condition is again $\nu =1,$ and the stimulus is here $+\delta \mathbf{\xi }
^{\nu }$ with $\nu $ changing every $1,5x10^{5}$ MCS. Thus, we conclude that
the switching phenomena is robust with respect to the type of pattern stored.

\section{Conclusion}

The set of equations (\ref{geme})--(\ref{pst}) provides a general framework
to model activity--dependent processes. Motivated by the behavior of
neurobiological systems, we adapted this to study the consequences of fast
noise acting on the synapses of an attractor neural network with a finite
number of stored patterns. We present in this paper two different scenarios
corresponding to noise distributions fulfilling (\ref{prod}) and (\ref{xp}),
respectively. In particular, assuming a local dependence on activity as in (\ref{bimod}), one obtains the local fields (\ref{locelf}), while a global
dependence as in (\ref{genbimod}) leads to (\ref{gloelf}). Under certain
assumptions, the system in the first of these cases is described by the
effective Hamiltonian (\ref{eh}). This reduces to a Hopfield system ---i.e.,
the familiar attractor neural network without any synaptic noise--- with
rescaled temperature and a threshold. This was already studied for a
Gaussian distribution of thresholds \cite{hertzB,threshold1,threshold2}.
Concerning stationary properties, this case is also similar to the one in \cite{bibitchkov}. 
A more intriguing behavior ensues when the noise
depends on the total presynaptic current arriving to the postsynaptic
neuron. We studied this case both analytically, by using a mean--field
hypothesis, and numerically by a series of Monte Carlo simulations using
single-neuron dynamics. The two approaches are
fully consistent with and complement each other.

Our model involves two main parameters. One is the \textit{temperature} $T$
which controls the stochastic evolution of the network activity. The other
parameter, $\Phi ,$ controls the depressing noise intensity. Varying this, the
system describes from normal operation to depression phenomena. A main
result is that the presynaptic noise induces the occurrence of a tricritical
point for certain values of these parameters, $(T_{c},\Phi _{c})=(1,-4/3).$
This separates (in the limit $\alpha \rightarrow 0$) first from second order
phase transitions between a retrieval phase and a non--retrieval phase.

The principal conclusion in this paper is that fast presynaptic \textit{noise
} may induce a nonequilibrium condition which results in an important
intensification of the network sensitivity to external stimulation. We
explicitly show that the noise may turn unstable the \textit{attractor} or
fixed point solution of the retrieval process, and the system then seeks for
another attractor. In particular, one observes switching from the stored
pattern to the corresponding antipattern for $M=1,$ and switching between
patterns for a larger number of stored patterns, $M.$ This behavior is most
interesting because it improves the network ability to detect changing
stimuli from the environment. We observe the switching to be very sensitive
to the forcing stimulus, but rather independent of the network initial state
or the thermal noise. It seems sensible to argue that, besides recognition,
the processes of class identification and categorization in nature might
follow a similar strategy. That is, different attractors may correspond to
different objects, and a dynamics conveniently perturbed by fast noise may
keep visiting the attractors belonging to a class which is characterized by
a certain degree of correlation between its elements \cite{cortesgs05}. In
fact, a similar mechanism seems at the basis of early olfactory processing
of insects \cite{olfact}, and instabilities of the same sort have been
described in the cortical activity of monkeys \cite{monkey} and other cases 
\cite{animals}.

Finally, we mention that the above complex behavior seems confirmed by
preliminary Monte Carlo simulations for a macroscopic number of stored
patterns, i.e., a finite loading parameter $\alpha =M/N\neq 0.$ On the other
hand, a mean--field approximation (see below) shows that the storage
capacity of the network is $\alpha _{c}=0.138,$ as in the Hopfield case \cite{amitAP}, 
for any $\Phi <0,$ while it is always smaller for $\Phi >0.$
This is in agreement with previous results concerning the effect of synaptic
depression in Hopfield--like systems \cite{torresCAPACITY,bibitchkov}. The
fact that a positive value of $\Phi $ tends to shallow the basin thus
destabilizing the attractor may be understood by a simple (mean--field)
argument which is confirmed by Monte Carlo simulations \cite{cortesgs05}.
Assume that the stationary activity shows just one overlap of order unity.
This corresponds to the \textit{condensed pattern;} the overlaps with the
rest, $M-1$ stored patterns is of order of $1/\sqrt{N}$ (\textit{non--condensed patterns}) \cite{hertzB}. The resulting probability of change
of the synaptic intensity, namely, $\left( 1+\alpha \right) \sum_{\nu
=1}^{P}(m^{\nu })^{2}$ is of order unity, and the local fields (\ref{gloelf}) follow as $h_{i}^{\mathrm{eff}}\sim -\Phi h_{i}^{\mathrm{Hopfield}}.$
Therefore, the storage capacity, which is computed at $T=0,$ is the same as
in the Hopfield case for any $\Phi <0,$ and always lower otherwise.

\section*{Acknowledgments}

We acknowledge financial support from MCyT--FEDER (project No. BFM2001-2841
and a \textit{Ram\'{o}n y Cajal} contract).


\begin{thebibliography}{}

\bibitem[Abbott and Kepler, 1990]{AK90}
Abbott, L.~F. and Kepler, T.~B. (1990).
\newblock Model neurons: From Hodgkin-Huxley to Hopfield.
\newblock {\em Lectures Notes in Physics}, 368,5--18.

\bibitem[Abbott and Regehr, 2004]{abb}
Abbott, L.~F. and Regehr, W.~G. (2004).
\newblock Synaptic computation.
\newblock {\em Nature}, 431,796--803.

\bibitem[Abeles et~al., 1995]{monkey}
Abeles, M., Bergman, H., Gat, I., Meilijson, I., Seidelman, E., Tishby, N., and
  Vaadia, E. (1995).
\newblock Cortical activity flips among quasi-stationary states.
\newblock {\em Proc. Natl. Acad. Sci. USA}, 92,8616--8620.

\bibitem[Allen and Stevens, 1994]{allenPNAS}
Allen, C. and Stevens, C.~F. (1994).
\newblock An evaluation of causes for unreliability of synaptic transmission.
\newblock {\em Proc. Natl. Acad. Sci. USA}, 91,10380--10383.

\bibitem[Amari, 1972]{amari72}
Amari, S. (1972).
\newblock Characteristics of random nets of analog neuron-like elements.
\newblock {\em IEEE Trans. Syst. Man. Cybern.}, 2,643--657.

\bibitem[Amit et~al., 1987]{amitAP}
Amit, D.~J., Gutfreund, H., and Sompolinsky, H. (1987).
\newblock Statistical mechanics of neural networks near saturation.
\newblock {\em Ann. Phys.}, 173,30--67.

\bibitem[Bibitchkov et~al., 2002]{bibitchkov}
Bibitchkov, D., Herrmann, J.~M., and Geisel, T. (2002).
\newblock Pattern storage and processing in attractor networks with short-time
  synaptic dynamics.
\newblock {\em Network: Comput. Neural Syst.}, 13,115--129.

\bibitem[Brunel and Hakim, 1999]{brunel99}
Brunel, N., and Hakim, V. (1999).
\newblock Fast Global Oscillations in Networks of Integrate-and-Fire
Neurons with Low Firing Rates.
\newblock {\em Neural Comp.}, 11,1621--1671.

\bibitem[Cortes et~al., 2005]{cortesgs05}
Cortes, J.~M., Garrido, P.~L., Kappen, H.~J., Marro, J., Morillas, C., Navidad,
  D., and Torres, J.~J. (2005).
\newblock Algorithms for identification and categorization.
\newblock {\em AIP Conf. Proc.}, In Press.

\bibitem[Cortes et~al., 2004]{cortes2T}
Cortes, J.~M., Garrido, P.~L., Marro, J., and Torres, J.~J. (2004).
\newblock Switching between memories in neural automata with synaptic noise.
\newblock {\em Neurocomputing}, 58-60,67--71.

\bibitem[Franks et~al., 2003]{fssjn03}
Franks, K.~M., Stevens, C.~F., and Sejnowski, T.~J. (2003).
\newblock Independent sources of quantal variability at single glutamatergic
  synapses.
\newblock {\em J. Neurosci.}, 23(8),3186--3195.

\bibitem[Gardiner, 2004]{gardinerB}
Gardiner, C.~W. (2004).
\newblock {\em Handbook of Stochastic Methods: for Physics, Chemistry and the
  Natural Sciences}.
\newblock Springer-Verlag.

\bibitem[Garrido and Marro, 1989]{garridoPRL}
Garrido, P.~L. and Marro, J. (1989).
\newblock Effective Hamiltonian description of nonequilibrium spin systems.
\newblock {\em Phys. Rev. Lett.}, 62,1929--1932.

\bibitem[Garrido and Marro, 1991]{GM91}
Garrido, P.~L. and Marro, J. (1991).
\newblock Nonequilibrium neural networks.
\newblock {\em Lecture Notes in Computer Science}, 540,25--32.

\bibitem[Garrido and Marro, 1994]{garridoJSP}
Garrido, P.~L. and Marro, J. (1994).
\newblock Kinetic lattice models of disorder.
\newblock {\em J. Stat. Phys.}, 74,663--686.

\bibitem[Garrido and Munoz, 1993]{garridoPRE}
Garrido, P.~L. and Munoz, M.~A. (1993).
\newblock Nonequilibrium lattice models: A case with effective Hamiltonian in d
  dimensions.
\newblock {\em Phys. Rev. E}, 48,R4153--R4155.

\bibitem[Hebb, 1949]{hebb}
Hebb, D.~O. (1949).
\newblock {\em The Organization of Behavior: A Neuropsychological Theory}.
\newblock Wiley.

\bibitem[Hertz et~al., 1991]{hertzB}
Hertz, J., Krogh, A., and Palmer, R. (1991).
\newblock {\em Introduction to the theory of neural computation}.
\newblock Addison-Wesley.

\bibitem[Hopfield, 1982]{hopfield}
Hopfield, J.~J. (1982).
\newblock Neural networks and physical systems with emergent collective
  computational abilities.
\newblock {\em Proc. Natl. Acad. Sci. USA}, 79,2554--2558.

\bibitem[Horn and Usher, 1989]{threshold1}
Horn, D. and Usher, M. (1989).
\newblock Neural networks with dynamical thresholds.
\newblock {\em Phys. Rev. A}, 40,1036--1044.

\bibitem[Lacomba and Marro, 1994]{lacombaEURLETT}
Lacomba, A. I.~L. and Marro, J. (1994).
\newblock Ising systems with conflicting dynamics: Exact results for random
  interactions and fields.
\newblock {\em Europhys. Lett.}, 25,169--174.

\bibitem[Laurent et~al., 2001]{olfact}
Laurent, G., Stopfer, M., Friedrich, R.~W., Rabinovich, M.~I., Volkovskii, A.,
  and Abarbanel, H. D.~I. (2001).
\newblock Odor encoding as an active, dynamical process: Experiments,
  computation and theory.
\newblock {\em Annu. Rev. Neurosci.}, 24,263--297.

\bibitem[Litinskii, 2002]{threshold2}
Litinskii, L.~B. (2002).
\newblock Hopfield model with a dynamic threshold.
\newblock {\em Theoretical and Mathematical Physics}, 130,136--151.

\bibitem[Marro and Dickman, 1999]{marroB}
Marro, J. and Dickman, R. (1999).
\newblock {\em Nonequilibrium Phase Transitions in Lattice Models}.
\newblock Cambridge University Press.

\bibitem[Marro et~al., 1999]{MTG99}
Marro, J., Torres, J.~J., and Garrido, P.~L. (1999).
\newblock Neural network in which synaptic patterns fluctuate with time.
\newblock {\em J. Stat. Phys.}, 94(1-6),837--858.

\bibitem[Miller and Schreiner, 2000]{animals}
Miller, L.~M. and Schreiner, C.~E. (2000).
\newblock Stimulus-based state control in the thalamocortical system.
\newblock {\em J. Neurosci.}, 20,7011--7016.

\bibitem[Pantic et~al., 2002]{torresNC}
Pantic, L., Torres, J.~J., Kappen, H.~J., and Gielen, S. C. A.~M. (2002).
\newblock Associative memmory with dynamic synapses.
\newblock {\em Neural Comp.}, 14,2903--2923.

\bibitem[Pitler and Alger, 1992]{PAJN92}
Pitler, T. and Alger, B.~E. (1992).
\newblock Postsynaptic spike firing reduces synaptic gaba(a) responses in
  hippocampal pyramidal cells.
\newblock {\em J. Neurosci.}, 12,4122--4132.

\bibitem[Sompolinsky, 1986]{S86}
Sompolinsky, H. (1986).
\newblock Neural networks with nonlinear synapses and a static noise.
\newblock {\em Phys. Rev. A}, 34,2571--2574.

\bibitem[Thomson et~al., 2002]{thom}
Thomson, A.~M., Bannister, A.~P., Mercer, A., and Morris, O.~T. (2002).
\newblock Target and temporal pattern selection at neocortical synapses.
\newblock {\em Philos. Trans. R. Soc. Lond. B Biol. Sci.}, 357,1781--1791.

\bibitem[Torres et~al., 1997]{torresPRA}
Torres, J.~J., Garrido, P.~L., and Marro, J. (1997).
\newblock Neural networks with fast time-variation of synapses.
\newblock {\em J. Phys. A: Math. Gen.}, 30,7801--7816.

\bibitem[Torres et~al., 2004]{torresNEUCOM}
Torres, J.~J., Munoz, M.~A., Marro, J., and Garrido, P.~L. (2004).
\newblock Influence of topology on the performance of a neural network.
\newblock {\em Neurocomputing}, 58-60,229--234.

\bibitem[Torres et~al., 2002]{torresCAPACITY}
Torres, J.~J., Pantic, L., and Kappen, H.~J. (2002).
\newblock Storage capacity of attractor neural networks with depressing
  synapses.
\newblock {\em Phys. Rev. E.}, 66,061910.

\bibitem[Tsodyks et~al., 1998]{tsodyksNC}
Tsodyks, M.~V., Pawelzik, K., and Markram, H. (1998).
\newblock Neural networks with dynamic synapses.
\newblock {\em Neural Comp.}, 10,821--835.

\bibitem[Zador, 1998]{zadorJN}
Zador, A. (1998).
\newblock Impact of synaptic unreliability on the information transmitted by
  spiking neurons.
\newblock {\em J. Neurophysiol.}, 79,1219--1229.

\end{thebibliography}
\end{document}